\documentclass[12pt]{article}
\usepackage{amsmath}
\usepackage{amssymb}
\usepackage[mathscr]{eucal}
\usepackage[usenames]{color}
\usepackage[utf8]{inputenc}
\usepackage{url}
\usepackage{longtable}
\usepackage{graphicx}
\usepackage{epstopdf}
\usepackage{subfigure}
\renewcommand{\arraystretch}{1.5}

\usepackage{url}
\usepackage[
bookmarks=true,
backref=true,
colorlinks=true,
linkcolor=red,
citecolor=blue,
urlcolor=green]{hyperref}

\newtheorem{theorem}{Theorem}

\newtheorem{remark}{Remark}

\numberwithin{equation}{section}

\definecolor{darkolivegreen}{rgb}{0.333333, 0.419608, 0.1843140}

\newcommand{\dy}{\partial_y}

\newcommand{\dt}{\partial_t}
\newcommand{\dx}{\partial_x}
\newcommand{\du}{\partial_u}

\usepackage{graphicx}
\usepackage{epstopdf}
\usepackage{subfigure}
\usepackage{color}
\usepackage{float}

\begin{document}

\title{\Large  Analysis of Lie symmetries and traveling wave solutions for the $(2+1)$-dimensional Boussinesq equation with general nonlinearity
}

\author{Şeyma Gönül\thanks{Corresponding author, gonul20@itu.edu.tr},  \; Cihangir \"{O}zemir\thanks{ozemir@itu.edu.tr}\\
\small Department of Mathematics, Faculty of Science and Letters,\\
\small Istanbul Technical University, 34467 Istanbul,
Türkiye}


\date{}

\maketitle

\begin{abstract}
In this study, we investigate Lie symmetries of the (2+1)-dimensional Boussinesq equation, which has been proposed to model the propagation of gravity waves on the water surface, with particular emphasis on the head-on collision of oblique waves. We consider this equation in a more general form involving an arbitrary function $f(u)$ and establish a complete Lie symmetry classification with respect to the admissible forms of the nonlinearity. 
For the canonical equations arising from the classification, we construct reductions to ordinary differential equations by using an optimal system of two-dimensional subalgebras.  Furthermore, we examine the exact solutions of the equation and analyze the stability of the traveling wave solutions.  

\vspace{0.5cm}
\end{abstract}

\section{Introduction}

In this article, we focus on the (2+1)-dimensional Boussinesq equation with a general nonlinearity

\begin{equation}\label{main}
    u_{tt} = u_{xx} + u_{yy}  + u_{xxxx}+ [f(u)]_{xx}. 
\end{equation}
We establish a complete Lie symmetry classification of Eq. \eqref{main}, determining the admissible forms of $f(u)$ with the corresponding Lie invariance algebras and reductions of the canonical equations to ordinary differential equations. We also find exact solutions for the quadratic and cubic nonlinearities and perform a phase plane analysis for the qualitative behaviour of traveling wave solutions.

Equation \eqref{main} represents  a more general form of 
\begin{equation}\label{maineq}
   u_{tt} +\alpha_1 u_{xx} + \alpha_2 u_{yy} + \beta(u^2)_{xx} +\alpha_3 u_{xxxx} = 0
\end{equation}
with an arbitrary function $f(u)$, appearing in Ref.~\cite{johnson} with the parameters $\alpha_1=\alpha_2=\alpha_3=-1$ and \(\beta = -3\), 
which was introduced to describe the propagation of gravity waves on the surface of water, in particular the head-on collision of oblique waves \cite{johnson}. Equation \eqref{maineq} is a significant model for describing the bidirectional propagation of long waves in shallow water under the influence of gravity, encompassing phenomena such as wave breaking, dispersion, water deformation, reflection, diffraction, refraction, wave-current interactions, and wave-induced tidal effects. 

Another motivation for this generalization comes from the equation
\begin{equation}\label{babaref}
u_{tt}=u_{xx}+u_{yy}+\left(6\alpha u u_x+6\beta u^2u_x+u_{xxx}\right)_x,
\end{equation}
which incorporates both quadratic and cubic nonlinearities and was studied in \cite{matsukawa1989soliton,matsukawa1988n}. The authors state that Eq.~\eqref{babaref} can be reduced to a Kadomtsev–Petviashvili (KP) type equation, which describes ion-acoustic waves in plasmas with negative ions propagating predominantly in the $x$-direction, with weak transverse variation in the y-direction.

By considering the nonlinearity as an arbitrary function rather than restricting it to a fixed polynomial structure, the model becomes theoretically broader and structurally more flexible. From a mathematical perspective, determining the Lie symmetry structure of the generalized equation is essential for understanding its reduction mechanisms, invariant structures, and possible integrable subclasses. A systematic classification with respect to $f(u)$ allows one to reveal how the functional form of the nonlinearity influences the symmetry algebra.

Numerous studies have explored this equation, yielding valuable insights. For instance, Allen and Rowlands  produced solitary wave solutions for the (2 + 1)-dimensional quadratic nonlinear Boussinesq equation \cite{allen1997transverse}. Ma et al. in \cite{ma} examined lump-type solutions using the Hirota bilinear method. In Ref. \cite{wang2024dynamics}, the authors developed novel resonant soliton solutions, such as $X$-shape and $Y$-shape solitons, for the (2+1)-dimensional Boussinesq equation using the Hirota bilinear approach, explored hybrid interactions and complex $N$-soliton solutions, and investigated traveling wave solutions, with the goal of enhancing understanding of the dynamics of shallow water waves. Also for Eq. \eqref{maineq}, Rogue wave solutions including single, triple and sixfold configurations were derived by Liu and Zhu \cite{zheng2022multiple}. Song and Shao  performed bifurcation analysis on the ordinary differential equations associated with its solitary wave solutions \cite{song2010exact}. In \cite{alam2015exact}, the exp($\phi(\eta)$)-expansion method has been used to derive exact traveling wave solutions for the (2+1)-dimensional Boussinesq equation, with solutions expressed in terms of exponential, hyperbolic, trigonometric, and rational functions. In \cite{moleleki2013solutions}, the simplest equation approach and Lie symmetry methods were applied to solve Eq. \eqref{maineq}. In addition, breathers and lumps of general order were discussed in \cite{liu2018general}. Soliton molecules and other traveling wave solutions were explored in \cite{wang2023soliton} and the soliton and breather solutions have been examined in references \cite{yi2008rational}, \cite{qiang2004exact}, \cite{wang2016characteristics}, while the mechanisms of breather transitions were addressed in \cite{gao2024breather}. The decomposition approach has also been applied to derive solutions in \cite{el2004decomposition}.
Hong et al. investigated the orbital stability of dn periodic wave solutions of the Boussinesq equation with quadratic-cubic nonlinear terms \cite{hong2023orbital}.
In \cite{helal2017stability}, the authors investigate the sixth-order nonlinear Boussinesq equation, deriving various soliton-like solutions using the extended auxiliary equation method and analyzing their stability.
In \cite{li2021analytical}, the global dynamics of the generalized Boussinesq equation with cubic nonlinearity and harmonic excitation were investigated, an equivalent wave equation was derived by means of perturbation methods, and persistent homoclinic structures exhibiting chaotic behavior were constructed.
Lian-Li Feng and colleagues studied the nonlocal symmetries and consistent Riccati expansion integrability of the (2+1)-dimensional Boussinesq equation, deriving its Painlevé expansion, Schwarzian form, and Bäcklund transformation, and analyzing the interaction solutions between solitons and cnoidal waves \cite{feng2020backlund}. 
In \cite{xu2016backlund}, the Bäcklund transformation, infinite conservation laws, and periodic wave solutions of Eq. \eqref{maineq} were explored. The (1+1) Boussinesq equation was found to have a three-dimensional Lie symmetry algebra by Levi and Winternitz \cite{levi1989non}.
		
		In \cite{bruzon2009travelling}, for the double-dispersion equation
		\begin{equation}
			u_{tt} = u_{xx} + a u_{xxtt} - b u_{xxxx} + d u_{xxt} + \left[f(u)\right]_{xx},
		\end{equation}
		Bruzon and Gandarias classified symmetry algebras. For the equation
		\begin{equation}
			u_{tt} - \Delta u - \Delta u_{tt} + \Delta^2 u - k\Delta u_t = \Delta f(u), \quad x \in \mathbb{R}^3,
		\end{equation}
		Yu et al.  classified  symmetry algebras and derived reduced equations \cite{yu2017lie}. Gandarias et al. \cite{gandarias2020conservation} analyzed the cases $n=1$ and $n=2$ for the equation
		\begin{equation}
			u_{tt} - u_{xx} + a u_{xxxx} - b u_{xxtt} = \left[f(u)\right]_{xx},
		\end{equation}
		and produced conservation laws. Recio et al. \cite{recio2016symmetries} studied the equation
		\begin{equation}
			u_{tt} = c u_{xx} + b u_{xxxx} + a u_{xxxxxx} + \left[f(u)\right]_{xx},
		\end{equation}
		and derived conservation laws.
		
		Meanwhile, Bruzon \cite{bruzon2009exact} examined the equation
		\begin{equation}
			u_{tt} = u_{xx} + u_{xxtt} - u_{xxxxtt} - c u_{xxxx} + \left[f(u)\right]_{xx},
		\end{equation}
		and obtained trigonometric, hyperbolic, and elliptic solutions via Lie symmetries.
		Dougalis et al. demonstrated the well-posedness of the initial value problem for a $(2+1)$-dimensional three-component Boussinesq-type system \cite{dougalis2007some}. Chen analyzed a similar system numerically, modeling the propagation of small-amplitude, long-wavelength surface waves in an incompressible, inviscid fluid \cite{chen2009numerical}. Dutykh and Dias derived more complex Boussinesq-type systems incorporating damping effects using asymptotic methods \cite{dutykh2007dissipative}. Zhao \cite{zhao2020three} examined the scalar Boussinesq equation
		\begin{equation}
			u_{tt} + \alpha_0 u_{yt} - \alpha_4 u_{yy} + \alpha_1 \varepsilon u_{xy} + \alpha_2 \varepsilon u_{xx}^2 + \alpha_3 \varepsilon^2 u_{xxxx} = 0,
		\end{equation}
		and obtained multi-soliton, multi-breather, and lump solutions using the Hirota bilinear method. Bruzon and Gandarias \cite{bruzon2009symmetries} analyzed the modified Boussinesq equation
		\begin{equation}
			u_{tt} = a u_{xx} + \left(u^{m+1}\right)_{xx} + b \left[u \left(u^m\right)_{xx}\right]_{xx},
		\end{equation}
		obtaining classical, non-classical, and potential symmetries as well as solutions in trigonometric, hyperbolic, exponential, and elliptic forms through reduced equations.
		Liu \cite{liu2000strong} analyzed the instability of solitary solutions and blow-up phenomena for the generalized Boussinesq equation
		\begin{equation}
			u_{tt} - u_{xx} + u_{xxxx} + \left[f(u)\right]_{xx} = 0.
		\end{equation}
In Ref. \cite{hussain2022lie}, the $(N+1)$-dimensional generalized nonlinear Boussinesq equation
\begin{equation}
u_{tt} = a u_{xx} + b (u^2)_{xx} + c u_{xxxx} + \sum_{j=1}^{N-1} d_j u_{y_j y_j},
\qquad b \neq 0
\end{equation}
is investigated within the framework of Lie symmetry analysis in order to derive similarity reductions, exact analytical solutions, and associated conservation laws. In \cite{farah2010global}, it is shown that the initial value problem for the cubic defocusing nonlinear Boussinesq equation
\begin{equation}
u_{tt} - u_{xx} + u_{xxxx} - (|u|^2 u)_{xx} = 0
\end{equation}
posed on the real line is globally well-posed in the Sobolev space $H^s(\mathbb{R})$ for $2/3 < s < 1$.

Matsukawa et al.\ first considered the modified Kadomtsev-Petviashvili equation
\begin{equation}
\left(u_t+8(u^3)_x+u_{xxx}\right)_x+u_{yy}=0,
\end{equation}
and the two-dimensional modified Boussinesq equation
\begin{equation}
u_{tt}=u_{xx}+u_{yy}+u_{xxxx}+8(u^3)_{xx},
\end{equation}
and obtained their \(N\)-soliton solutions by the Hirota bilinear method \cite{matsukawa1988n}. They showed that the former admits only parallel two-soliton propagation, whereas the latter allows two-soliton solutions propagating in different directions under a certain condition. In a subsequent work, they generalized these models to a Kadomtsev-Petviashvili type equation and also to the two-dimensional Boussinesq-type equation
\eqref{babaref},
which takes into account both quadratic and cubic nonlinearities \cite{matsukawa1989soliton}.  Moreover, they pointed out that, for \(\alpha=0\) and \(\beta=4\), the generalized equations reduce exactly to the modified equations studied in their earlier paper \cite{matsukawa1988n} .
       
          In this work, we consider the $(2+1)$-dimensional Boussinesq equation \eqref{main}, which can be used to describe the propagation of gravity waves on water surfaces in its more generalized form involving an arbitrary function $f(u)$,
            where $x$  represents the direction of wave propagation, $y$ represents the weak dependence, and $f(u)$ is an arbitrary function  \cite{johnson}. Eq.~\eqref{maineq} was originally derived in Ref. \cite{johnson} by introducing a weak dependence on the second spatial dimension into the classical Boussinesq equation.

In contrast to the existing literature, where the (2+1)-dimensional Boussinesq equation is mostly studied under fixed quadratic or cubic nonlinearities, we consider the generalized form involving an arbitrary nonlinear function $f(u)$. This formulation allows a complete Lie symmetry classification with respect to the admissible forms of the nonlinearity and reveals richer symmetry structures beyond the standard polynomial cases.
From a physical perspective, the nonlinear term is directly related to the properties of the  medium of propagation. Therefore, determining which forms of $f(u)$ admit enhanced symmetry structures also provides information about physically relevant cases compatible with the intrinsic symmetries of the model. Furthermore, we do not restrict the analysis to symmetry classification only. For the important quadratic–cubic case, exact traveling-wave solutions are derived in elliptic, hyperbolic, and trigonometric forms, and the associated reduced dynamical system is analyzed through phase portraits and stability theory. This unified approach clarifies the relation between symmetry structure, nonlinear wave patterns, and qualitative dynamical behavior.

This paper is organized as follows. In Section 2, the Lie symmetry algebra of the equation is determined with respect to different forms of $f(u)$, and reductions of the canonical equations to ordinary differential equations are obtained by using an optimal system of two-dimensional subalgebras. In Section 3, traveling-wave solutions of the Boussinesq equation containing cubic--quadratic nonlinear terms are presented. Finally, in Section 4, the stability of the traveling-wave solutions is analyzed in order to understand the dynamical behavior of the solutions.

\section{Lie Algebra and Canonical Equations}

	We analyze the (2+1)-dimensional Boussinesq equation \eqref{main} with arbitrary nonlinearity using Lie group methods, within the theoretical framework presented in \cite{Bluman, Hydon, Stephanie}.
\subsection{Lie algebra}
Eq. \eqref{main} will be invariant under the transformations generated by the vector field
	\begin{equation}
		V=\tau(t,x,y,u)\dt+\psi (t,x,y,u)\dx+\rho (t,x,y,u)\dy+\mu  (t,x,y,u) \du
	\end{equation}
	where \begin{subequations}
	\begin{eqnarray}
  \mu_{tt} - \mu_{yy} - \mu_{xx} - f_{u}\mu_{xx} - \mu_{xxxx}= 0,& \\
\tau_y - \rho_t=0, & \\
 \rho_y - 2 \psi_x =0, & \\
 3 \psi_{xx}- 2 \mu_{xu}  =0, & \\
\frac{1}{2}\mu f_{uu} + \psi_x +  f_{u}\psi_x + 3 \mu_{xxu} - 2 \psi_{xxx} =0,&  \\
  2 \psi_x-  \tau_t = 0,& \\
 \rho_{yy} - \rho_{tt}- 2 \mu_{yu} =0,& \\
 \psi_{xx}+ \psi_{xxxx} + f_{u}\psi_{xx}- 2 f_{uu}\big(\mu_x +\mu_{xu}\big)  - 4 \mu_{xxxu}  =0,& \\
 \mu f_{uuu} + f_{uu}\mu_u + 2 f_{uu}\psi_x =0,&\\
\tau_{yy} - \tau_{tt} + 2 \mu_{tu} =0,&  \\
\tau_x =\tau_u=
\psi_t =\psi_y =   
\psi_u = 
\rho_x=
\rho_u =
\mu_{uu} = 
0. & 
\end{eqnarray}\end{subequations}
After solving the determining equations we obtain
	\begin{equation}
	\begin{aligned}
						\tau(t,x,y,u)&=k_3+k_2 t+k_4 y,\\
		\psi (t,x,y,u)&=k_0+ \frac{ k_2}{2}x,\\
		\rho (t,x,y,u)&=k_5+k_4 t+k_2 y,
	\end{aligned}
	\end{equation}
and
	\begin{equation}
		\mu(t,x,y,u)=k_1 u+\xi(t,y)
	\end{equation}	
	where $k_0$, $k_1$, $k_2$, $k_3$, $k_4$, and $k_5$ are arbitrary constants, while $\xi(t,y)$ satisfies
\begin{subequations}\label{xi_eqs}
	\begin{eqnarray}
		 \xi_{tt}- \xi_{yy}  &=& 0, \\
		 k_2 + k_2 f_u + k_1 u f_{uu} + \xi f_{uu} &=& 0.
	\end{eqnarray}
\end{subequations}	
We assume that $f_{uu} \neq 0$. From Eqs.~(\ref{xi_eqs}a,b), we can clearly see that $\xi=\mu_0$ where $\mu_0$ is a constant; therefore,
	\begin{equation}
			\label{first} k_2 + k_2 f_u + k_1 u f_{uu} + \mu_0 f_{uu} = 0.
\end{equation}
  Depending on the cases of the constants given here, we will obtain $f$ in various forms. We can present these forms as follows: 
\begin{subequations}\label{formf1}
    \begin{eqnarray}
         &&\text{(A)} \quad f(u)=\alpha e^{\beta u}+\gamma u+\delta,  \label{fA}\\
         && \text{(B)} \quad f(u)=\alpha \ln|u+\beta_0|+\gamma u+\delta, \label{fB}\\
         && \text{(C)} \quad f(u)=\alpha( u+\beta_0)^n+\gamma  u+\delta, \quad n\neq 0,1,  \label{fC}\\
          && \text{(D)} \quad f(u)=\alpha  (u+\beta_0) \ln|u+\beta_0|+\gamma u+ \delta,  \label{fD}
      \end{eqnarray} 
\end{subequations}
where $\alpha,\beta,\beta_0,\gamma,n$ are arbitrary constants and  $\alpha\neq0$, $\beta\neq0$. The value of $\delta$ is not important in equation \eqref{main}. Let us introduce the  transformations
\begin{subequations}
    \begin{eqnarray}
         &&T_1 : \quad   u = \alpha \bar  u-\beta_0,  \quad \bar x =x, \quad  \bar y= y,  \quad \bar t= t,  \\
         &&T_2 : \quad   u =   \bar u-\beta_0,  \quad \bar x =  x, \quad  \bar y=y,   \quad \bar t=t.
      \end{eqnarray} 
\end{subequations}
When we apply the transformation $T_1$ to Eq. \eqref{main} with $f$ in \eqref{fB}, and when we apply the transformation $T_2$ to Eq. \eqref{main} with $f$ as  in \eqref{fC},  \eqref{fD} and drop the bars in the variables, the forms of the nonlinearity $f(u)$ simplify to  
\begin{subequations}\label{formf2}
    \begin{eqnarray}
         &&\text{(A)} \quad f(u)=\alpha e^{\beta u} +\gamma u, \\
         &&\text{(B)} \quad f(u)=\ln|u| +\gamma  u ,\\
         &&\text{(C)} \quad f(u)=\alpha u^n+\gamma  u, \quad n\neq 0,1,\\
          &&\text{(D)} \quad f(u)=\alpha u\ln|u|+\gamma u.
      \end{eqnarray} 
\end{subequations}

Let us introduce the vector fields
\begin{equation}\label{principal}
X_1=\dt,  \quad X_2=\dx, \quad X_3=\dy,\quad  X_4=y \dt+t \dy
\end{equation}
and also 
\begin{align}
X_{5.A}&=t \dt+\frac{x}{2}\dx+y \dy-\frac{1}{\beta} \du,\\
X_{5.B}&=t \dt+\frac{x}{2}\dx+y \dy+u \du,\\
X_{5.C_1}&=t \dt+\frac{x}{2}\dx+y \dy+\frac{u}{1-n} \du, \quad n\neq 0,1,2,\\
X_{5.C_2}&=t \dt+\frac{x}{2}\dx+y \dy-\Big(u+\frac{1+\gamma}{2\alpha}\Big)\du.
\end{align}
\begin{theorem}\label{teorem1}
The Lie algebra of the (2+1)- dimensional Boussinesq equation \eqref{main} is classified in the following cases.
\begin{itemize}
\item[(i)] For an arbitrary $f(u)$, the Lie algebra of Eq. \eqref{main} is 4-dimensional and  has the basis
\begin{equation}\label{principal2}
L_1= \langle X_1, X_2, X_3, X_4\rangle.
\end{equation}
This includes the  forms of f given in Case (A), (B), (C) with $\gamma \neq -1$ and Case (D). 
\item[(ii)] When $f(u)=\alpha e^{\beta u} - u$, the Lie algebra is 5-dimensional and is spanned by 
\begin{equation}
L_2=\langle X_1, X_2, X_3, X_4,X_{5.A}\rangle.
\end{equation}
\item[(iii)] When $f(u)=\ln| u|  - u$, the Lie algebra is 5-dimensional and is spanned by 
\begin{equation}
L_3=\langle X_1, X_2, X_3, X_4,X_{5.B}\rangle.
\end{equation}
\item[(iv)] When $f(u)=\alpha u^n - u$, $n\neq 0,1,2$, the Lie algebra is 5-dimensional and is spanned by 
\begin{equation}
L_4=\langle X_1, X_2, X_3, X_4,X_{5.C_{1}}\rangle.
\end{equation}
\item[(v)] When $f(u)=\alpha u^2 +\gamma u$, the Lie algebra is 5-dimensional and is spanned by 
\begin{equation}
L_5=\langle X_1, X_2, X_3, X_4,X_{5.C_{2}}\rangle.
\end{equation}
\end{itemize}
Therefore, the Lie algebra of Eq. \eqref{main} is at most 5-dimensional.
\end{theorem}

\subsection{Canonical classes of equations}
Below we list each of the equations mentioned in Theorem 1 and present the commutation relations between the generators of the symmetry algebra.
\subsubsection{$\mathbf{f(u)=\alpha e^{\beta u}-u}$}
In this case,
Eq. \eqref{main} takes the form
	\begin{equation}\label{eqexp}
			u_{tt} =  u_{yy} + u_{xxxx} + \alpha\left(  e^{\beta u} \right)_{xx}.
		\end{equation}
        The following is an account of the nonzero commutation relations:
\begin{equation}\label{KA}
   [X_1,X_4]=X_3 ,\quad [X_1 ,X_{5.A}]= X_1,\quad [X_2 ,X_{5.A}]=\frac{1}{2}X_2, \quad [X_3,X_4]=X_1, \quad [X_3,X_{5.A}]=X_3.
\end{equation}
\subsubsection{$\mathbf{f(u)=ln|u|-u}$}
In this case,
Eq. \eqref{main} takes the form
	\begin{equation}\label{eqlog}
			u_{tt} =  u_{yy} + u_{xxxx} + \big(\ln|u| \big)_{xx}.
		\end{equation}
The nonzero commutation relations are given as follows:
\begin{equation}\label{KB}
  [X_1,X_4]=X_3, \quad [X_1,X_{5.B}]= X_1,\quad [X_2,X_{5.B}]=\frac{1}{2}X_2, \quad [X_3,X_4]=X_1, \quad [X_3,X_{5.B}]=X_3.
\end{equation}

\subsubsection{$\mathbf{f(u)=\alpha u^n-u},  \quad n\neq 0,1,2.$ }
In this case,
Eq. \eqref{main} takes the form
	\begin{equation}\label{equn}
			u_{tt} =  u_{yy} + u_{xxxx} + \alpha\left( u^n\right)_{xx}.
		\end{equation}        
The nonzero commutation relations are given as follows:
\begin{equation}\label{KC1}
   [X_1,X_4]=X_3, \quad [X_1,X_{5.C_{1}}]= X_1,\quad [X_2,X_{5.C_{1}} ]=\frac{1}{2}X_2, \quad  [X_3,X_4]=X_1, \quad [X_3,X_{5.C_{1}} ]=X_3.
\end{equation}
\subsubsection{$\mathbf{f(u)=\alpha u^2+\gamma u}$}
In this case,
Eq. \eqref{main}  is of  the form
	\begin{equation}\label{equ2}
			u_{tt} = (1+\gamma)u_{xx}+ u_{yy} + u_{xxxx} + \alpha\left( u^2\right)_{xx}.
		\end{equation}       
The nonzero commutation relations are as follows:
\begin{equation}\label{KC2}
 [X_1,X_4]=X_3, \quad [X_1,X_{5.C_{2}}]= X_1, \quad [X_2,X_{5.C_{2}} ]=\frac{1}{2}X_2, \quad [X_3,X_4]=X_1,  \quad [X_3,X_{5.C_{2}} ]=X_3.
\end{equation}
\begin{remark}
According to Theorem 1 part (i), for any form of $f(u)$, the generalized Boussinesq equation \eqref{main} admits the 4-dimensional Lie algebra with generators given in \eqref{principal}. For some specific forms of the nonlinearity, the dimension of the Lie algebra extends to 5. These forms are specified in parts (ii)-(v) of Theorem 1.

It is apparent that the available literature on the (2+1)-dimensional Boussinesq equation is quite limited and the functional forms of the nonlinearity $f(u)$ investigated are restricted to  quadratic and cubic type nonlinearities.  Lie symmetry analysis performed in this section covers these power type nonlinearities. Besides these usual forms considered in literature, our analysis identifies other forms of the nonlinearities that can lead to a richer symmetry structure. The exponential and logarithmic nonlinearities presented in parts (ii) and (iii) of ~Theorem 1 lead to equations \eqref{eqexp} and \eqref{eqlog} with a five-dimensional Lie algebra.
\end{remark}
\subsection{Reductions to ODEs}
In this section we are going to obtain the reductions of the canonical equations we presented above. For an arbitrary $f(u)$, Eq. \eqref{main} admits the 4-dimensional Lie algebra $L_1$. The  canonical equations in Eqs. \eqref{eqexp}, \eqref{eqlog}, \eqref{equn} and  \eqref{equ2}  admit the 5-dimensional Lie algebras $L_2$,  $L_3$,  $L_4$,  $L_5$ which are isomoporphic as they have the same commutation relations. In order to find the reductions of these equations to ODEs, we need an optimal system of two-dimensional subalgebras of this 5-dimensional algebra. We obtained this optimal system by using the package SymboLie, for which we refer to  \cite{amata2023automatic,amata2024optimal, amata2025symbolic}. This optimal system is composed of  eight two-dimensional subalgebras $S_i$, $i=1,\ldots,8$ listed below. 
\begin{table}[h]
\centering
\centering
\caption{Subalgebras and reduced equations for Case (i). $f(u)$ is arbitrary. $c_0$ and  $c_1$ are arbitrary real numbers. } 
\label{table:1}
\renewcommand{\arraystretch}{2}
\begin{tabular}{|l|l|}
\hline
\multicolumn{2}{|c|}{
  $u_{tt} = u_{xx} + u_{yy}  + u_{xxxx}+ [f(u)]_{xx}$ } \\
\hline
\textbf{Subalgebra} & \textbf{Reduced ODE or Solution} \\
\hline
$S_1$ 
& $u = M(x)$ \\
& $M'' + M + f(M) = c_1 x + c_0$\\
\hline
$S_2$ 
& $u = c_1 \ln|y^2 - t^2| + c_0$ \\
\hline
$S_3$  
& $\alpha_1=1$: \quad $u=M(y-t)$,  $M$: arbitrary \\
&$\alpha_1  =-1 $: \quad  $u=y+ t+c_0$ \\
\hline
$S_4$ 
& 
$u = M(\eta), \quad \eta = \alpha_2 x - y + \alpha_1 t$ \\
 &$M'' + M +  f(M) = c_1 \eta + c_0$ \\
\hline

\end{tabular}
\end{table}

\begin{table}[h]
\small
\centering
\caption{Subalgebras and solutions for Case (ii). $c_0$ and $c_1$ are arbitrary real numbers.}
\label{table:1}
\renewcommand{\arraystretch}{2}
\begin{tabular}{|l|l|}
\hline
\multicolumn{2}{|c|}{
$u_{tt} =  u_{yy} + u_{xxxx} + \alpha\left(  e^{\beta u} \right)_{xx} $} \\
\hline
\textbf{Subalgebra} & \textbf{Reduced ODE or Solution} \\
\hline 
$\langle X_2,X_{5.A} \rangle$   &$u  =M(\eta)-\dfrac{1}{\beta} \ln |t| , \quad \eta = y/t$  \\
 &$ \beta(1-\eta^2)M''-2\beta \eta M'-1 =0$
\\
\hline
$\langle X_2, X_4+\alpha_3 X_{5.A}\rangle$ 
& 
$\alpha_3\neq \mp 1$: \quad 
$u =\left(c_1(\alpha_3-1)-\dfrac{\alpha_3}{\beta(\alpha_3+1)}\right)\ln|y+t|
-
c_1(\alpha_3+1)\ln|y-t|+c_0$
\\
& 
$\alpha_3=1$: \qquad 
$u= M(\eta)
-
\dfrac{1}{2\beta}\ln|y+t|, \quad \eta=y-t,
\quad M\text{: arbitrary}$
\\
&
$\alpha_3=-1$: \qquad 
$u = 
M(\eta)
-
\dfrac{1}{2\beta}\ln|y-t|, \quad \eta=y+t,
\quad M\text{: arbitrary}$
\\
\hline
$\langle X_4,X_{5.A} \rangle$  
& 
$u = M(\eta)-\dfrac{1}{2\beta}\ln (y^2-t^2), \quad \eta=x^2/\sqrt{y^2-t^2} $
\\
& $16\eta^2 M^{(4)}
+48\eta M^{(3)}
+(12+\eta^2)M''
+\eta M'
+2\alpha\beta e^{\beta M}
\left[
M'
+2\beta\eta\bigl(M'\bigr)^2
+2\eta M''
\right]
=0$ 
\\
\hline
$\langle X_1+\alpha_1 X_3,X_{5.A} \rangle$ 
&$u =
M(\eta)
-\dfrac{1}{\beta}\ln(y-\alpha_1 t),
\quad
\eta=x^2/(y-\alpha_1 t)$\\
& $
8\eta^2 M^{(4)}
+
24\eta M^{(3)}
+
6 M'' 
+
\alpha\beta e^{\beta M}
\left[
M'
+
2\beta\eta\bigl(M'\bigr)^2
+
2\eta M''
\right]
=0
$ \\
\hline
\end{tabular}
\end{table}

\begin{table}[h]
\centering
\caption{Subalgebras and solutions for Case (iii). $c_0$ and $c_1$ are arbitrary real numbers.}
\label{table:1}
\renewcommand{\arraystretch}{2}
\begin{tabular}{|l|l|}
\hline
\multicolumn{2}{|c|}{
\centering $u_{tt} = u_{yy} + u_{xxxx} + \left( \ln |u|\right)_{xx} $} \\
\hline

\textbf{Subalgebra} & \textbf{Reduced ODE or Solution} \\
\hline 
$\langle X_2,X_{5.B} \rangle$   &$u  =c_1 y+c_0 t$  \\
\hline
$\langle X_2, X_4+\alpha_3 X_{5.B}\rangle$ 
& 
$\alpha_3\neq \mp 1$: \quad 
$u = c_0|y-t|^{\frac{\alpha_3}{\alpha_3-1}}
+c_1|y+t|^{\frac{\alpha_3}{\alpha_3+1}}$
\\
& 
$\alpha_3=1$: \qquad 
$u = c_1|y+t|^{\frac{1}{2}}$
\\
&
$\alpha_3=-1$: \qquad 
$u = 
c_0|y-t|^{\frac{1}{2}}$
\\
\hline
$\langle X_4,X_{5.B} \rangle$  
& 
$u = (y^2-t^2)^{\frac{1}{2}}M(\eta), \quad \eta = x^2   / \sqrt{y^2-t^2}$
\\
&$16\eta^2 M^2 M^{(4)}
+48\eta M^2 M^{(3)}
+\left(12+\eta^2\right)M^2 M''
+4\eta M M'' $ \\
&$
-4\eta (M')^2
-\eta M^2 M'
+2M M'
+M^3=0 $
\\
\hline
$\langle X_1+\alpha_1 X_3,X_{5.B} \rangle$ 
&$u = (y-\alpha_1t)M(\eta), \quad \eta = x^2 / (y-\alpha_1 t)$\\
&$8\eta^2 M^2 M^{(4)}
+24\eta M^2 M^{(3)}
+6 M^2 M'' $ \\
&
$+2\eta M M''
-2\eta (M')^2
+M M'
=0 $ \\
\hline
\end{tabular}
\end{table}

\begin{table}[h]
\centering
\caption{Subalgebras and solutions for Case (iv). $c_0$ and $c_1$ are arbitrary real numbers.}
\label{table:1}
\renewcommand{\arraystretch}{2}
\begin{tabular}{|l|l|}
\hline
\multicolumn{2}{|c|}{
$u_{tt} = u_{yy} + u_{xxxx} + \alpha\left( u^n\right)_{xx}, \quad n\neq 0,1,2 $} \\
\hline
\textbf{Subalgebra} & \textbf{Reduced ODE or Solution} \\
\hline 
$\langle X_2,X_{5.C_1} \rangle$   &$u  =t^{\frac{1}{1-n}}M(\eta), \quad \eta = y/t$  \\
 &$(n-1)^2(\eta^2-1)M''+2n(n-1)\eta M'+nM=0$
\\
\hline
$\langle X_2, X_4+\alpha_3 X_{5.C_1}\rangle$ 
& 
$\alpha_3\neq \mp 1$: \quad 
$u = c_0|y-t|^{\frac{\alpha_3}{(1-n)(\alpha_3-1)}}
+c_1|y+t|^{\frac{\alpha_3}{(1-n)(\alpha_3+1)}}$
\\
& 
$\alpha_3=1$: \qquad 
$u = c_1|y+t|^{\frac{1}{2(1-n)}}$
\\
&
$\alpha_3=-1$: \qquad 
$u = 
c_0|y-t|^{\frac{1}{2(1-n)}}$
\\
\hline
$\langle X_4,X_{5.C_1} \rangle$  
& 
$u = (y^2-t^2)^{\frac{1}{2(1-n)}}M(\eta), \quad \eta = x^2   / \sqrt{y^2-t^2}$
\\
& $(n-1)^2 \Big[ 16\eta^2 M^{(4)}
+48\eta M^{(3)}
+(12+\eta^2)M''$  \\
&$
+4n\alpha \eta M^{\,n-1}M''
+4n\alpha (n-1)\eta M^{\,n-2}(M')^2$ \\
&$
+2n\alpha M^{\,n-1}M' \Big ]
+(n^2-1)\eta M'
+M=0 $
\\
\hline
$\langle X_1+\alpha_1 X_3,X_{5.C_1} \rangle$ 
&$u = (y-\alpha_1t)^{1/(1-n)}M(\eta), \quad \eta = x^2 / (y-\alpha_1 t)$\\

&$(n-1)^2\Big[ 16\eta^2 M^{(4)}
+48\eta M^{(3)}
+12 M''$ \\
&
$+4n\alpha\eta M^{\,n-1}M''
+4n\alpha(n-1)\eta M^{\,n-2}(M')^2
+2n\alpha M^{\,n-1}M' \Big]
=0 .$
 \\
\hline
\end{tabular}
\end{table}

\begin{table}[h]
\footnotesize
\centering
\caption{Subalgebras and solutions for Case (v). $c_0$ and $c_1$ are arbitrary real numbers.}
\label{table:1}
\renewcommand{\arraystretch}{2}
\begin{tabular}{|l|l|}
\hline
\multicolumn{2}{|c|}{
$u_{tt} = (1+\gamma)u_{xx}+ u_{yy} + u_{xxxx} + \alpha\left( u^2\right)_{xx} $} \\
\hline
\textbf{Subalgebra} & \textbf{Reduced ODE or Solution} \\
\hline 
$\langle X_2,X_{5.C_2} \rangle$   &$u  = -\dfrac{1+\gamma}{2\alpha}+\dfrac{1}{t}M(\eta), \quad \eta = y/t$  \\
 &$(\eta^2-1)M''+4\eta M'+2M=0$
\\
\hline
$\langle X_2, X_4+\alpha_3 X_{5.C_2}\rangle$ 
& 
$\alpha_3\neq \mp 1$: \quad  
$u = -\dfrac{1+\gamma}{2\alpha}
+c_0|y-t|^{\frac{\alpha_3}{1-\alpha_3}}
+c_1|y+t|^{\frac{-\alpha_3}{1+\alpha_3}}$
\\
& 
$\alpha_3=1$: \quad
$u = -\dfrac{1+\gamma}{2\alpha}
+c_1|y+t|^{-1/2}$
\\
&
$\alpha_3=-1$: \quad
$u = -\dfrac{1+\gamma}{2\alpha}
+c_1|y-t|^{-1/2}$ \rule[-3ex]{0pt}{6ex}
\\
\hline
$\langle X_4,X_{5.C_2} \rangle$  
& 
$u =-\dfrac{1+\gamma}{2\alpha}+ (y^2-t^2)^{-1/2}M(\eta), \quad \eta = x^2   / \sqrt{y^2-t^2}$
\\
&
$16\eta^2 M^{(4)}
+ 48\eta M'''
+ (12
+ \eta^2) M''
+ 8\alpha \eta M M'' 
+ M
+ 3\eta M'
+ 4\alpha M M' 
+ 8\alpha \eta (M')^2=0$
\\
\hline
$\langle X_1+\alpha_1 X_3,X_{5.C_2} \rangle$ 
&$u = -\dfrac{1+\gamma}{2\alpha}+\dfrac{1}{y-\alpha_1t}M(\eta), \quad \eta = x^2 / (y-\alpha_1 t)$\\
& $
4\eta^2 M^{(4)}
+ 12\eta M^{(3)}
+ 3 M''
+\alpha M M' +2\alpha \eta M M'' 
+ 2\alpha \eta (M')^2=0
$ \\
\hline
\end{tabular}
\end{table}

\begin{align}
&S_1=\langle X_1,X_3 \rangle        &S_5&=\langle X_2,X_5 \rangle \\
&S_2=\langle X_2,X_4 \rangle        &S_6&=\langle X_2, X_4+\alpha_3 X_5\rangle \\
&S_3=\langle X_1+\alpha_1 X_3,X_2 \rangle        &S_7&=\langle X_4,X_5 \rangle \\
&S_4=\langle X_1+\alpha_1 X_3, X_2+\alpha_2 X_3 \rangle        &S_8&=\langle X_1+\alpha_1 X_3,X_5 \rangle 
\end{align}

Here we denoted the generators $X_{5.A}$, $X_{5.B}$, $X_{5.C_1}$ and  $X_{5.C_2}$  as $X_5$.  Also,
\begin{equation}
\alpha_1^2= 1, \quad  \alpha_2^2 = 1, \quad   \alpha_3\neq 0.
\end{equation}

As the  subalgebras $S_1$, $S_2$, $S_3$, $S_4$ involve only the generators $X_1$, $X_2$, $X_3$ and $X_4$, the reductions obtained through these 2-dimensional subalgebras will be valid for arbitrary $f(u)$. These reductions were listed in Table 1. The two-dimensional subalgebras $S_5$, $S_6$, $S_7$ and  $S_8$ involve the infinitesimal generator $X_5$, which differs slightly with respect to the nonlinearity. In Table 2, we present the similarity variables and reduced ODEs for Eq.  \eqref{eqexp}. Table 3 presents the results for Eq. \eqref{eqlog}, Table 4 for Eq. \eqref{equn}, and Table 5 for Eq. \eqref{equ2}. In some of the cases, the reduced equations were straightforward to solve. In those cases, we expressed the solution.

Having presented our main results on Lie symmetry analysis, we now compare them with the results obtained in \cite{gandarias3} for the double dispersion equation
\begin{equation}\label{doubled}
u_{tt}=u_{xx}+au_{ttxx}+bu_{xxxx}+du_{txx}+(f(u))_{xx}+\beta u_{yy},
\end{equation}
which includes the family \eqref{main} as a particular case. 

Introducing the potential transformation $u=v_x$, the authors of \cite{gandarias3} rewrite Eq.~\eqref{doubled} in the form
\begin{equation}\label{doubledv}
v_{tt}=v_{xx}+av_{ttxx}+bv_{xxxx}+dv_{txx}+(f(v_x))_{x}+\beta v_{yy}.
\end{equation}
They then determine the Lie symmetries of Eq.~\eqref{doubledv} for the admissible forms of $f(v_x)$. 

The main difference between Eq.~\eqref{doubledv} with $a=d=0$ and Eq.~\eqref{main} lies in the structure of the nonlinearity represented by $f$. Despite this distinction, the admissible functional forms of $f(v_x)$ obtained in \cite{gandarias3} coincide with the nonlinearities found in the present work. However, this difference becomes apparent in the structure of the corresponding Lie algebras.

First, according to our Theorem~1, the Lie algebra associated with Eq.~\eqref{main} is at most five-dimensional. In contrast, for any form of  $f(v_x)$, Eq.~\eqref{doubledv} admits an additional infinite-dimensional abelian Lie algebra generated by
\begin{equation}
\big\langle g(y+\sqrt{\beta}\,t)\partial_v,\,
h(y-\sqrt{\beta}\,t)\partial_v \big\rangle,
\end{equation}
where $g$ and $h$ are arbitrary functions of a single variable. No such infinite-dimensional component appears in the Lie algebra of Eq.~\eqref{main}. This discrepancy originates from the passage from Eq.~\eqref{doubled} to Eq.~\eqref{doubledv} through the potential transformation $u=v_x$.

Comparing the symmetry generators obtained for specific nonlinearities, we observe that, in the power-law and logarithmic cases, our results parallel the finite-dimensional parts of the Lie algebras presented in cases \emph{(vii)} and \emph{(viii)} of Theorem~2 in \cite{gandarias2}. On the other hand, our results are different in the case of exponential nonlinearity. We also differ by  the special quadratic case. Let us note that similar arguments can be made for \cite{gandarias1}, which studies the power case nonlinearity for the potential form equation.

Furthermore, we also perfom a complete symmetry reduction analysis.   Overall, this comparison indicates that, reporting directly on equation \eqref{main}, the present work both extends and complements the symmetry classification results available in \cite{gandarias2}. 

\section{Exact Solutions}

In this section, we will concentrate on the analytical solutions of \eqref{mainexact} that arise in the case $f(u)=\alpha u^2+\beta u^3$ where  $\alpha$ and $\beta $ are constants, that is, we  consider the equation
\begin{equation}\label{mainexact}
u_{tt} =  u_{xx}+u_{yy} + u_{xxxx} + \left( \alpha u^2+\beta u^3 \right)_{xx}.
\end{equation}
The main motivation for studying exact solutions in this case is provided by Ref.~\cite{matsukawa1988n} and Ref.~\cite{matsukawa1989soliton}. 
We shall concentrate on  exact solutions of  traveling wave type
\begin{equation}\label{ansatz}
u(x,y,t)=U(\xi),\qquad \xi=kx+my-ct
\end{equation}
where $k$, $m$ and $c$ are real constants.
Remember that 
$
X_1=\partial_t$,  $X_2=\partial_x$, $X_3=\partial_y
$
are symmetries for \eqref{main} for any $f$. The solutions of \eqref{main} which are invariant under the transformations generated by the two-dimensional abelian subalgebra   
$
S_4=\left\langle X_1+\alpha_1X_3,\; X_2+\alpha_2X_3\right\rangle 
$
has the form
\begin{equation}\label{eta}
u(x,y,t)=M(\eta),\qquad \eta=\alpha_2x-y+\alpha_1 t.
\end{equation}
The reduced ODE for $M$ is given in the last  line of Table 1. If one considers the two-dimensional abelian subalgebra   $\tilde S_4=\langle mX_1+cX_3, m X_2-kX_3\rangle$ which is isomorphic to $S_4$,  the general traveling wave ansatz \eqref{ansatz}
is obtained. Obviously, $\eta$ is a special case of  \eqref{ansatz} with $k=\alpha_2$, $m=-1$, $c=-\alpha_1$.
 Thus, the traveling wave solutions studied in this section
belong to the class of invariant solutions generated by the element 
$S_4$ of the optimal system of two-dimensional subelgabras.

To obtain exact solutions of  \eqref{mainexact} we employ the ansatz
\eqref{ansatz} and get
\begin{equation}
k^4 U^{(4)}+(k^2+m^2-c^2)U''+k^2(\alpha U^2+\beta U^3)''=0. 
\end{equation}
If we integrate twice, we get 
\begin{equation}\label{dinamikgecis}
k^4 U''+(k^2+m^2-c^2)U+\alpha k^2 U^2+\beta k^2 U^3=K 
\end{equation}
where the constant in the first integration was chosen as zero.
 Multiplying both sides of the equation by $U' $, we obtain
\begin{equation}\label{four}
   \left( \frac{dU}{d \xi} \right )^2=A_4 U^4+A_3 U^3+A_2 U^2+A_1 U+ A_0=\Psi (U)
\end{equation}
where
\begin{equation}
    A_4=-\frac{ \beta}{2 k^2},\quad A_3=-\frac{2 \alpha}{ 3 k^2},\quad A_2=\frac{c^2-k^2-m^2}{k^4},\quad A_1=\frac{2 K }{k^4}
\end{equation}
and $A_0$ is an arbitrary constant. Assume that  $U_1$, $U_2$, $U_3$ and $U_4$ are zeros of $\Psi(U)$.        Here, our solutions will depend on the signs of the coefficients of the polynomial $\Psi$.       
Eq. \eqref{four} can be written as
\begin{equation}\label{exactint}
   \frac{dU}{\sqrt{A_4 U^4+A_3 U^3+A_2 U^2+A_1 U+ A_0}} =\varepsilon d \xi,
\end{equation}
where $\varepsilon=\mp1$.

\noindent \textbf{Case I.} 
    In this case we introduce two different types of elliptic solutions. \\ 
    
   \noindent \textbf{(i)} Here we assume that $A_1$, $A_2$, $A_3$ and $A_4$ are different from zero; that is, we consider the case when the polynomial $\Psi(U)$ has four distinct real roots.  
		Let us first write
		\begin{equation}
			\frac{d U}{\sqrt{A_4 (U-U_1)(U-U_2)(U-U_3)(U-U_4)}}= \varepsilon d \xi,
		\end{equation}
		assuming $A_4>0$. Let us consider the case $U_1> U_2 >U \geq U_3 >U_4 $. There are other intervals in this case, but we will just examine the one we have chosen for $U$. If we use the results from the handbook \cite{handbook} for this interval,  
		we obtain
     		\begin{equation}
			\int_{U}^{U_2}\frac{d  \tau}{\sqrt{A_4 (U_1-\tau)(U_2-\tau)(\tau-U_3)(\tau-U_4)}}=\frac{ p_0}{\sqrt{A_4}} \, \mathrm{sn}^{-1}\Big(\sqrt{\frac{(U_1 -U_3) (U_2 -U)}{(U_2 -U_3)(U_1 -U)}},\kappa \Big)
		\end{equation}
via the transformation
		\begin{equation}
        \mathrm{sn}^2 (w)=\frac{(U_1-U_3) (U_2-\tau)}{(U_2 -U_3)(U_1-\tau)}, 
        \end{equation}
        where
		\begin{equation}
			  p_0=\frac{2}{\sqrt{(U_1 -U_3)(U_2 -U_4)}}, \quad
			\kappa^2=\frac{(U_2 -U_3) (U_1 -U_4)}{(U_1 -U_3)(U_2 -U_4)}.
		\end{equation}
		This gives us the elliptic function solution of $U$ as 
		\begin{equation}
			\label{sn}U(\xi)=\frac{U_2-U_1 E \mathrm{sn}^{2}\left( \varepsilon{\frac{\sqrt{A_4}}{ p_0}}(\xi+\delta_0),\kappa\right) }{1-E \mathrm{sn}^2\left( \varepsilon \frac{\sqrt{A_4}}{p_0 }(\xi+\delta_0),\kappa\right) }
		\end{equation}
		where $0< E=\dfrac{(U_2-U_3)}{(U_1-U_3)}< \kappa^2$.\\
        
\begin{figure*}[hbt!] 
 	\centering
 	{\label{main:a}\includegraphics[scale=.8]{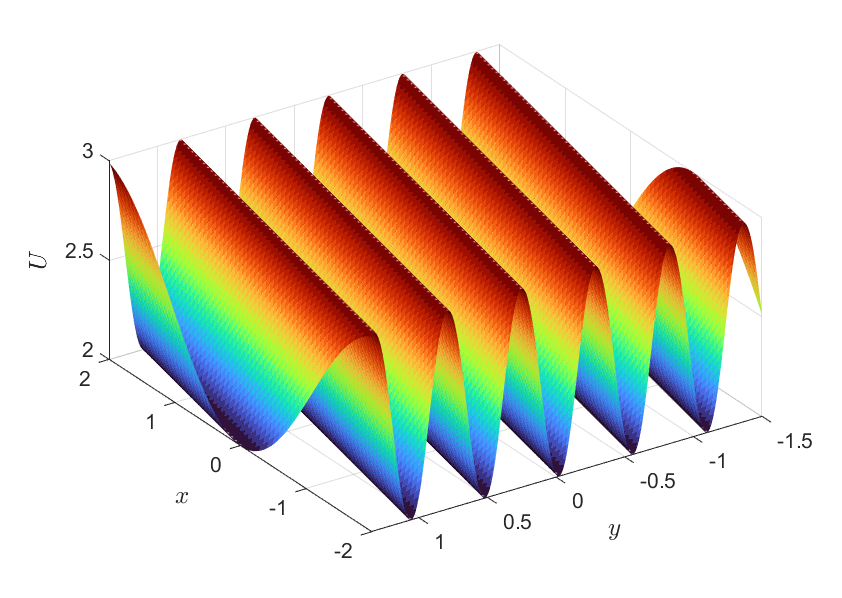}}
 	\caption{ Numerical simulation of the   solution \eqref{sn} for $t=0$. The values of parameters are
 		$k=1,\
 		m=8$,\
 		$c=10,\
 		\alpha=15,\ \beta=-2, \
 		K=-25,\
 		A_0=24$ and the roots of the Eq.~\eqref{sn} are $4,3,2,1$, respectively.}
 	\label{snf}
 \end{figure*}

     \noindent For the selected parameter values \(k=1\), \(m=8\), \(c=10\), \(\alpha=15\), \(\beta=-2\), \(K=-25\), and \(A_0=24\), with roots \(U_1=4\), \(U_2=3\), \(U_3=2\), and \(U_4=1\), the solution \eqref{sn} takes for \(\delta_0=0\) the form
\begin{equation}
u(x,y,t)=
\frac{6-4\,\mathrm{sn}^2\!\left(x+8y-10t,\frac{\sqrt{3}}{2}\right)}
{2-\mathrm{sn}^2\!\left(x+8y-10t,\frac{\sqrt{3}}{2}\right)}.
\end{equation}
This solution is illustrated in Fig.~\ref{snf} for $t=0$.
\\
\noindent \textbf{(ii) }Now let us assume  $A_4<0$   and  $\Psi(U)$ has two different, real zeros $U_1$, $U_2$ and two complex conjugate zeros $U_3$, $U_4$. Then, we can express \eqref{exactint} in the form 
\begin{equation}
\frac{dU}{\sqrt{(U_1-U)(U-U_2)(U-s)(U-\bar s)}}=\varepsilon \sqrt{-A_4} d\xi, \qquad \varepsilon=\mp 1.
\end{equation}
We consider the interval $ U_1 > U > U_2 $. Using the results from \cite{handbook}, 
the left hand side is integrated to 
\begin{equation}
\int_{U_2}^{U}\frac{d\tau}{\sqrt{(U_1-\tau)(\tau-U_2)(\tau-s)(\tau-\bar s)}}=g \, \mathrm{cn}^{-1}(\cos\psi,\kappa)
\end{equation}
via the transformation 
\begin{equation}
\mathrm{cn}(w)=\frac{B(U_1-\tau)-A(\tau-U_2)}{B(U_1-\tau)+A(\tau-U_2)}
\end{equation}
where 
\begin{equation}
g=\frac{1}{\sqrt{AB}}, \quad \cos \psi =\frac{B(U_1-U)-A(U-U_2)}{B(U_1-U)+A(U-U_2)}, \quad \kappa^2=\frac{(U_1-U_2)^2-(A-B)^2}{4AB}
\end{equation}
and
\begin{equation}
 A^2=(U_1-b_1)^2+a_1^2, \quad B^2=(U_2 -b_1)^2+a_1^2, \quad b_1=\frac{s+\bar s}{2}, \quad a_1^2=-\frac{(s-\bar s)^2}{4}.
\end{equation}
 Therefore one gets
\begin{equation}
U(\xi)= \frac{BU_1+AU_2+(AU_2-BU_1) ~\mathrm{cn}\left[\frac{ \sqrt{-A_4}}{g}(\xi-\xi_0),\kappa\right]}{A+B+(A-B)  ~\mathrm{cn}\left[\frac{ \sqrt{-A_4}}{g}(\xi-\xi_0),\kappa\right]}.
\end{equation}


	 \noindent	\textbf{Case II.} Next, we set $A_3 = 0$ and $A_1 = 0$. We emphasize that the solutions obtained in this case correspond to $A_3 = 0$, i.e., to a purely cubic nonlinearity with $\alpha = 0$. Under these assumptions, Eq.~\eqref{four} reduces to
        \begin{equation}\label{power}
             (U')^2=A_4 U^4+ A_2 U^2+A_0.
        \end{equation}
		If we define $\Omega=U^2$,  the discriminant of the equation $A_4 \Omega ^2+A_2 \Omega+A_0=0 $ is $ \Delta=A_2 ^2-4 A_4 A_0$. If $ \Delta=0$, then \eqref{power} takes the form
		\begin{equation}
			\frac{dU}{2A_4U^2+A_2}=\frac{\varepsilon}{2\sqrt{A_4}}\ d\xi
		\end{equation}
		where $\varepsilon =\mp 1$.\\

\noindent \textbf{(i)} \  If $A_2>0$, we find
		\begin{equation}
			U(\xi)=\varepsilon\sqrt{\frac{A_2}{2A_4}}\tan\big( \sqrt{\frac{A_2}{2}}(\xi+\delta_0)\big)
		\end{equation}
		where $A_4$ and $A_0$ have the same sign, $\delta_0$ is integration constant.	

\begin{figure*}[h!]	
	\centering
	{\includegraphics[scale=.7]{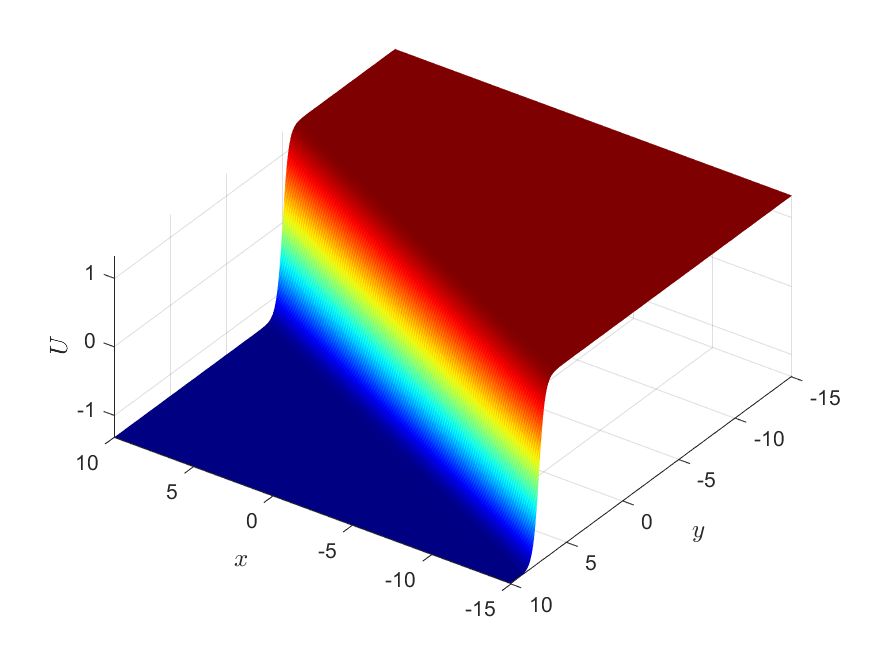}}
	\caption{Numerical simulation of the   solution \eqref{tanh} for $t=0$. The values of parameters are  $k=1$,
		$m=1$,
		$c=0.5$, $\beta=-1$ and $A_0=1.53125$. }
	\label{tanhf}
    
\end{figure*}

\noindent \textbf{(ii)} \  If $A_2<0$, we obtain that
		\begin{align}\label{tanh}
			&U(\xi)=-\varepsilon\sqrt{\frac{-A_2}{2A_4}}\tanh \Big[\sqrt{-\frac{A_2}{2}} (\xi+\delta_0)\Big].
		\end{align}		
        To illustrate the solution \eqref{tanh}, we take \(k=m=1\), \(c=0.5\), \(\beta=-1\), and \(A_0=1.53125\). Then, for  \(\delta_0=0\) it reduces to
\begin{equation}
u(x,y,t)=\frac{\sqrt{7}}{2}\tanh\left(\sqrt{\frac{7}{8}}\,(x+y-0.5t)\right).
\end{equation}
The corresponding profiles of this solution are presented in Fig.~\ref{tanhf} for \(t=0\). 

\noindent Particularly, when $A_0=0$ in \eqref{power},  the equation reduces to
		\begin{equation}
			\frac{d U}{U \sqrt{A_4 U^2+A_2}} = \varepsilon d \xi
		\end{equation}
		where $\varepsilon =\mp 1$. We can evaluate this integral in the following three different cases.

\begin{figure*}[h!]		
	\centering 
	\subfigure[]{\label{main:a}\includegraphics[scale=.60]{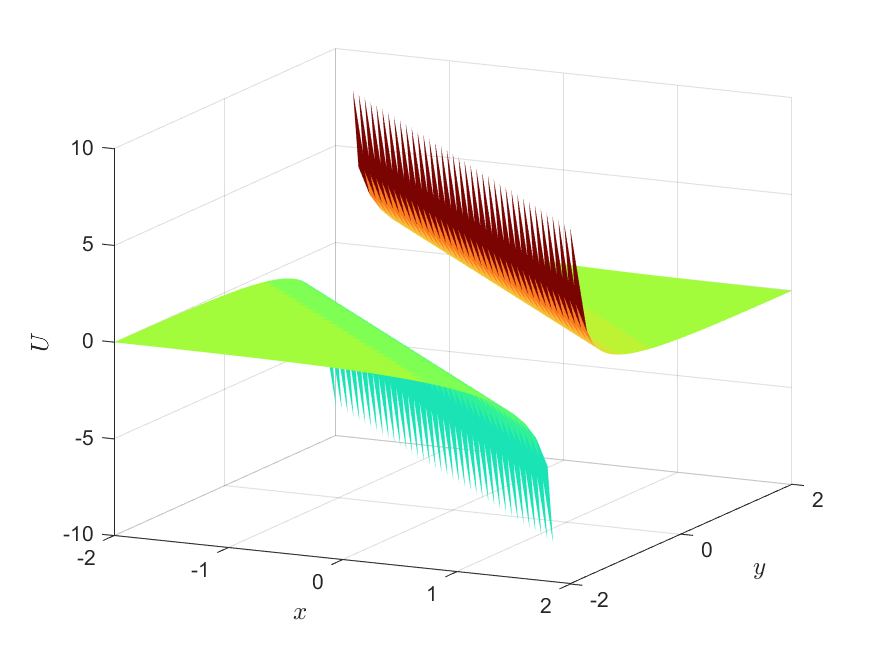}}
	\subfigure[]{\label{main:b}\includegraphics[scale=.50]{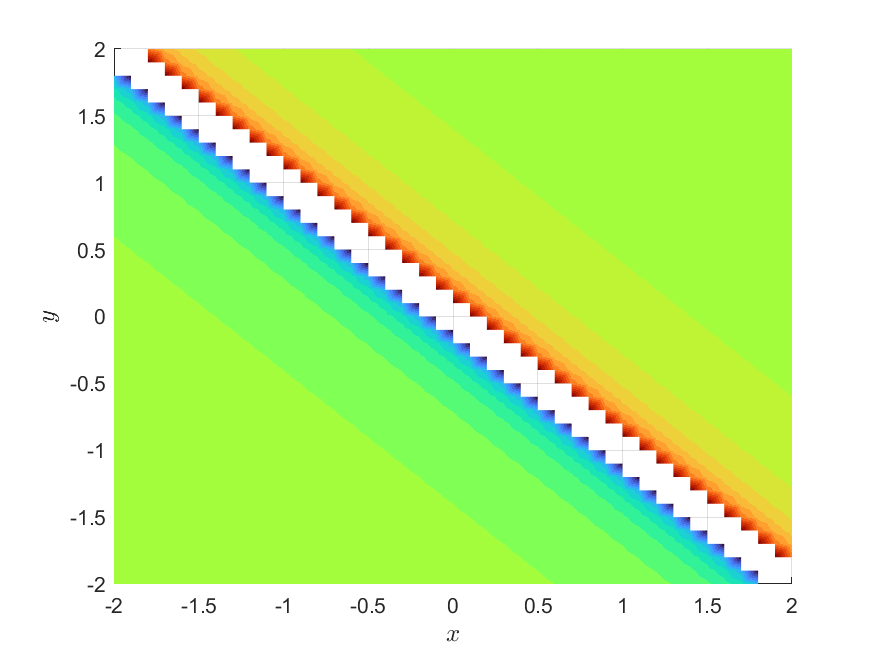}}
	
	\caption{Numerical simulations of the solution \eqref{csch} for $t=0$. The values of parameters are $k=1$,
		$m=1$,
		$c=1.5$ and $\beta=-1$.}
	\label{cschf}
\end{figure*}

		\noindent \textbf{(iii)} \ In case $A_4>0$ and $A_2>0$, we get
		\begin{align} \label{csch}
			U(\xi)&=\varepsilon\sqrt{\frac{A_2}{A_4}}\ \mathrm{csch}\Big(\sqrt{A_2}(\xi+\delta_0)\Big).
		\end{align}
        For \(k=m=1\), \(c=1.5\), and \(\beta=-1\), the solution \eqref{csch}  takes for \(\delta_0=0\) the form
\begin{equation} \label{cschn}
u(x,y,t)=\varepsilon \frac{1}{\sqrt2}\,\mathrm{csch}\!\left(\frac{1}{2}(x+y-1.5t)\right).
\end{equation}
The corresponding profiles of this solution presented in Fig.~\ref{cschf} for $t=0$.

\begin{figure*}[h!]		
	\centering 
	\subfigure[]{\label{main:a}\includegraphics[scale=.60]{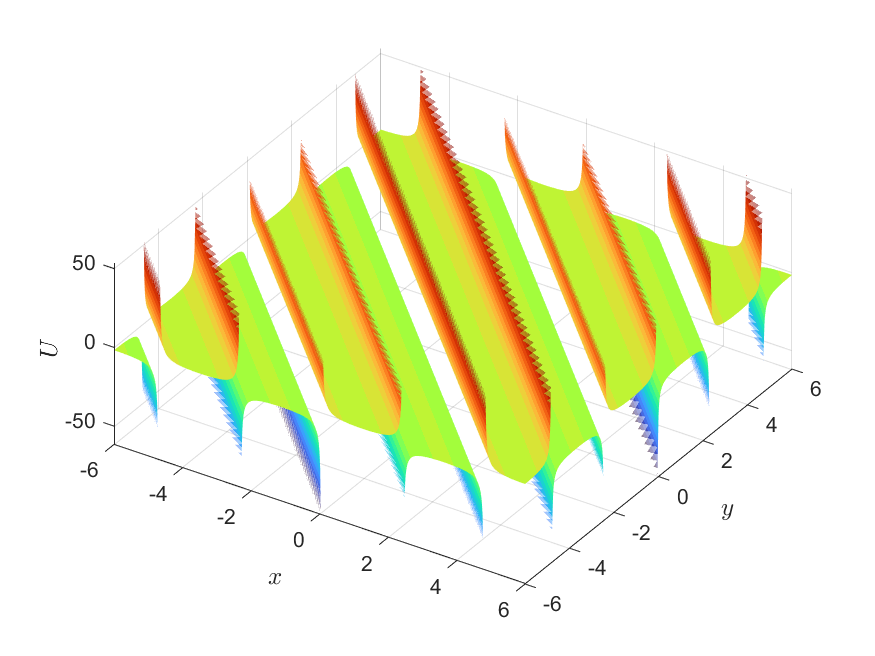}}
	\subfigure[]{\label{main:b}\includegraphics[scale=.50]{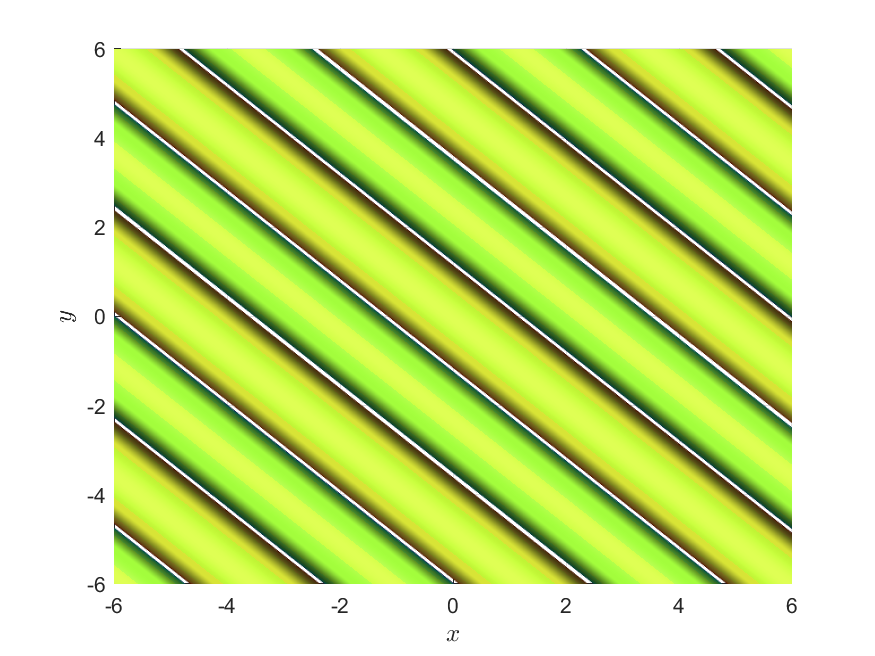}}
	
	\caption{Numerical simulations of the solution \eqref{sec} for $t=0$. The values of parameters are $k=1$,
		$m=1$,
		$c=0.5$ and $\beta=-1$.}
	\label{secf}
\end{figure*}

\noindent		\textbf{(iv)} \ In case $A_4>0$ and $A_2<0$, we obtain
		\begin{align}\label{sec}
			U(\xi)&=\sqrt{-\frac{A_2}{A_4}}\ \mathrm{sec}\Big(\sqrt{-A_2}(\xi+\delta_0)\Big).	
		\end{align}
        With the choice \(k=m=1\), \(c=0.5\), \(\beta=-1\), and \(\delta_0 =0\) the solution \eqref{sec} reduces to
\begin{equation}\label{secn}
u(x,y,t)=\sqrt{\frac72}\,\mathrm{sec}\!\left(\frac{\sqrt7}{2}(x+y-0.5t)\right)
.
\end{equation}
The corresponding profiles of this solution presented in Fig.~\ref{secf} for $t=0$.

\noindent \textbf{(v)} \ In case $A_4<0$ and $A_2>0$, we get
		\begin{align}	\label{sech}
		 U(\xi) &=\sqrt{-\frac{A_2}{A_4}}\ \mathrm{sech}\Big(\sqrt{A_2}(\xi+\delta_0)\Big).			
		\end{align}
Taking \(k=m=1\), \(c=2\), and \(\beta=1\),  the solution \eqref{sech} takes for \(\delta_0=0\) the form 
\begin{equation}\label{sechn}
u(x,y,t)=2\,\mathrm{sech} ~ \left(\sqrt2~ (x+y-2t) \right).  
\end{equation}

\begin{figure*}[h!]		
	\centering
	\subfigure[]{\label{main:a}\includegraphics[scale=.7]{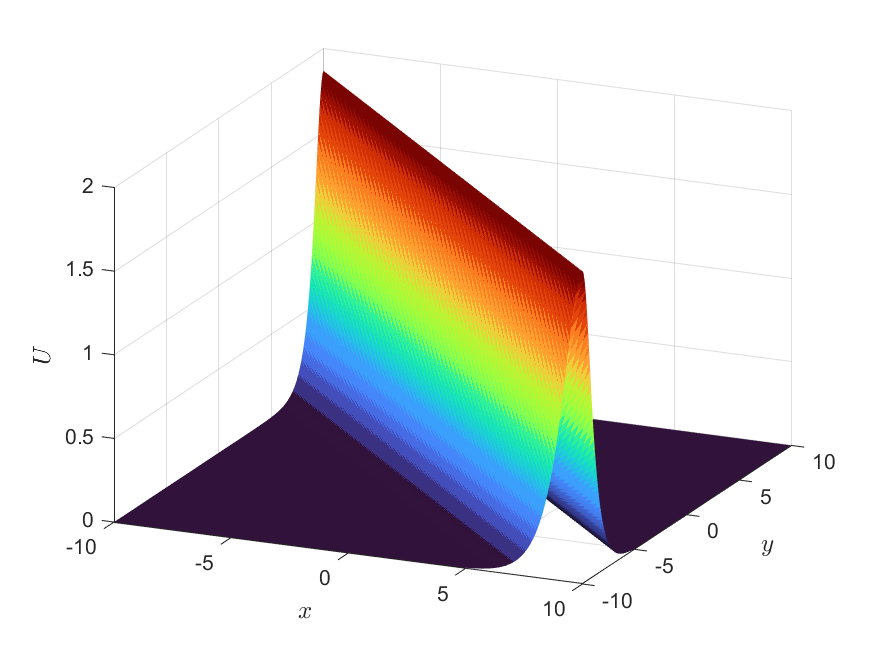}}
	
	\caption{Numerical simulation of the solution \eqref{sech} for $t=0$. The values of parameters are $k=1,\
		m=1$,\
		$c=2$ and $\beta=1$.}
	\label{sechf}
\end{figure*}

\noindent The corresponding profiles of solutions  \eqref{sechn} are presented in Fig.~\ref{sechf} for \(t=0\).
\section{Stability Analysis}

The dynamical system approach is widely used in the analysis of nonlinear partial differential equations because it reveals the qualitative structure of solutions. In this work, phase portrait classification is employed to investigate the dynamical behavior of the reduced equation. Phase analysis provides an effective framework for understanding nonlinear systems, particularly when explicit analytical solutions are not available. Since solution behavior may change significantly with parameter variations, examining the phase diagram is essential for identifying periodic, solitary, and other qualitative solution structures.

The second-order ordinary differential equation in \eqref{dinamikgecis} is rewritten as the following two-dimensional first-order dynamical system by taking $K=0$ and $\alpha=0$:
\begin{equation}
\begin{cases} \label{dinamiksistem}
U' = y,\\
y' = 2 A_4 U^3 + A_2 U,
\end{cases}
\end{equation}
where
\begin{equation}
A_4 = -\frac{\beta}{2 k^4}, 
\qquad
A_2 = -\frac{k^2+m^2-c^2}{k^4}.
\end{equation}
 See that the case $\alpha=0$ corresponds to a pure cubic nonlinearity in \eqref{mainexact}. Let us state that a discussion for the stability analysis of  \eqref{dinamiksistem} was performed in   \cite{han2023chaotic}. In this subsection we include this analysis for completeness and express connections with the analytical solutions we obtained above.

 Suppose $U(\xi)$ is a solution of system \eqref{dinamiksistem} satisfying
\begin{equation}
\lim_{\xi\to -\infty}U(\xi)=u_1,
\qquad
\lim_{\xi\to +\infty}U(\xi)=u_2	
\end{equation}
where $u_1$ and $u_2$ are constants. When $u_1=u_2$,  $\big(U(\xi),U'(\xi)\big)$ represents a homoclinic orbit and the original equation admits a solitary wave solution. When $u_1\neq u_2$, the trajectory represents a heteroclinic orbit, leading to kink or antikink wave solutions depending on the ordering of $u_1$ and $u_2$. Furthermore, if system \eqref{dinamiksistem} possesses closed phase trajectories, then the reduced equation admits periodic solutions \cite{han2023chaotic}.\\ 
System \eqref{dinamiksistem} is a Hamiltonian system. Then, the Hamiltonian function can be written as
\begin{equation}
H(U,y)=\frac12 y^2-\left(\frac{A_4}{2}U^4+\frac{A_2}{2}U^2\right)=h,
\label{eq:Hamiltonian}
\end{equation}
where $\frac12 y^2$ represents the kinetic energy and
$-\dfrac{A_4}{2}U^4-\dfrac{A_2}{2}U^2$ corresponds to the potential energy.
Solving $U'=y'=0$ yields the equilibrium points of system \eqref{dinamiksistem}.
Then we have
\begin{equation}
y=0, \ \ U(2 A_4U^2+A_2)=0.
\end{equation}
Thus, the system always possesses the equilibrium point $(0,0)$. Additional equilibrium points exist when $-A_2/A_4>0$, and are given by
\begin{equation}
\left(\pm\sqrt{-\frac{A_2}{2 A_4}},\,0\right).
\end{equation}

\noindent To classify the equilibria, we consider the Jacobian matrix
\begin{equation}
J(U,y)=
\begin{pmatrix}
0 & 1\\
6A_4U^2+A_2 & 0
\end{pmatrix}.
\label{eq:Jac}
\end{equation}
The characteristic equation is
\[
\lambda^2=6A_4U^2+A_2.
\]
Hence, the equilibrium $(U,0)$ is a saddle point when
$6A_4U^2+A_2>0$, a center when
$6A_4U^2+A_2<0$, and a degenerate (cusp-type) when the expression vanishes. In particular, for $(0,0)$ the classification depends on the sign of $A_{2}$.

We now examine how the phase portraits change according to the signs of $A_4$ and $A_2$, which leads to four distinct cases.

\begin{figure}[t]
\centering

\subfigure[$A_4>0$, $A_2<0$]{ \label{d_a}
\includegraphics[width=0.45\textwidth]{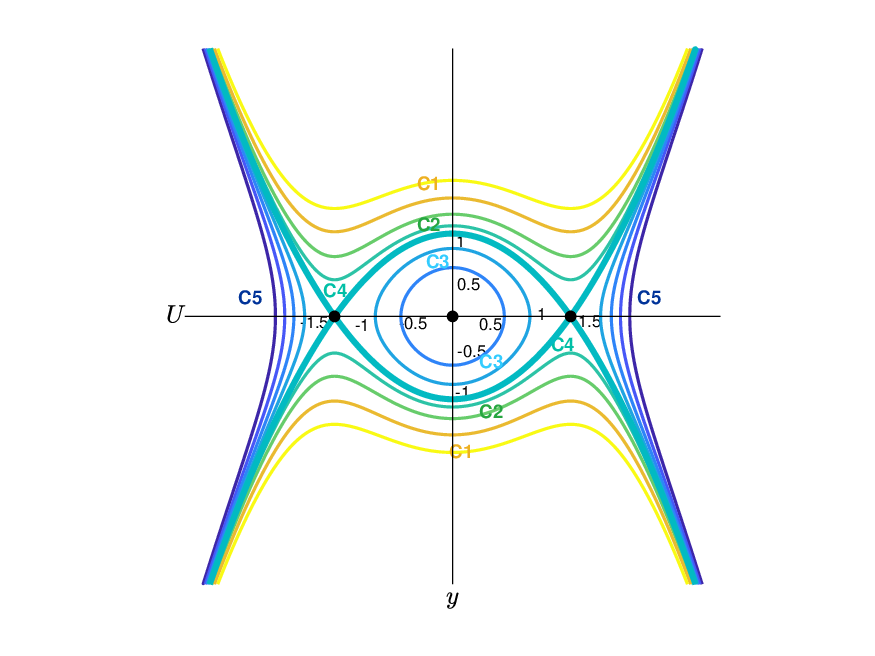}
}
\hfill
\subfigure[$A_4>0$, $A_2>0$]{\label{d_b}
\includegraphics[width=0.45\textwidth]{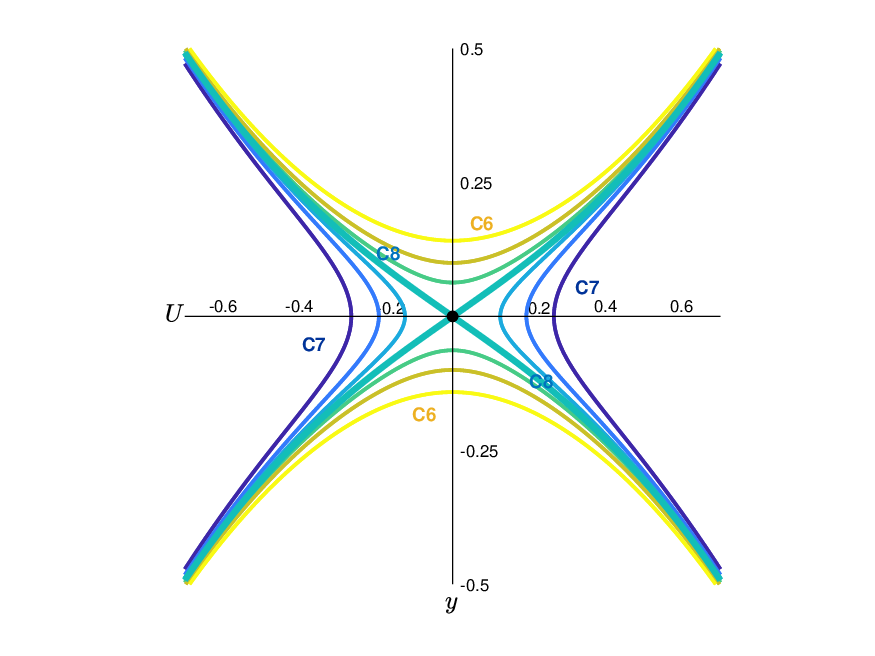}
}

\vspace{0.3cm}

\subfigure[$A_4<0$, $A_2<0$]{\label{d_c}
\includegraphics[width=0.45\textwidth]{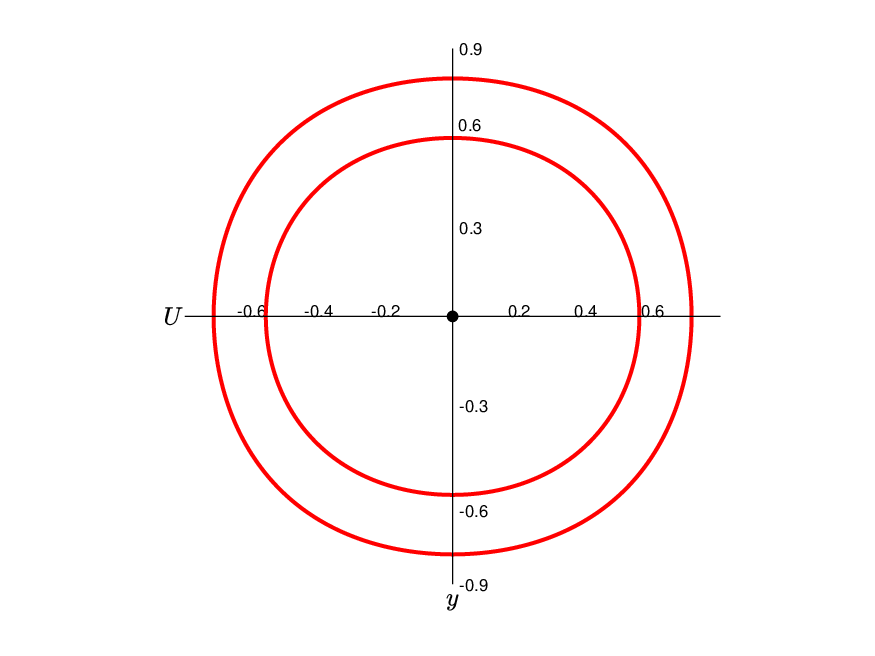}
}
\hfill
\subfigure[$A_4<0$, $A_2>0$]{\label{d_d}
\includegraphics[width=0.45\textwidth]{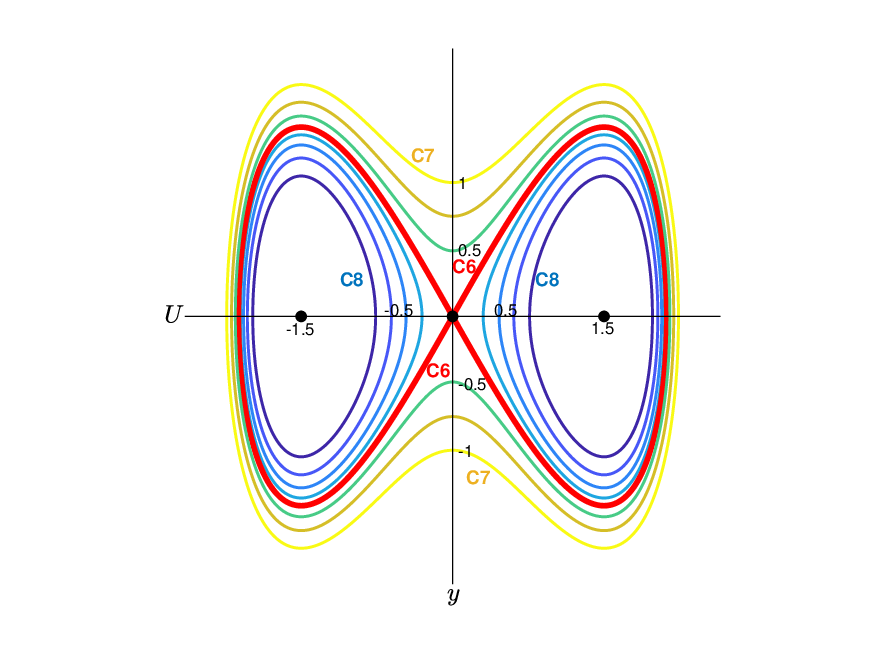}
}
\caption{Phase portraits associated with the four parameter regimes. For each case, representative parameter values are chosen as (a) $k=1$, $m=1$, $c=0.5$, $\beta=-1$; (b) $k=1$, $m=1$, $c=1.5$, $\beta=-1$; (c) $k=1$, $m=1$, $c=1$, $\beta=1$; (d) $k=1$, $m=1$, $c=2$, $\beta=1$.  }
\end{figure}

\paragraph{Case 1: $A_4>0,\;A_2<0$.}
In this case, the system has three equilibrium points: the origin $(0,0)$, which is a center, and two symmetric saddle points
\[
\left(\pm\sqrt{-\frac{A_2}{2 A_4}},0\right).
\]
The phase portraits are illustrated in Fig.~\ref{d_a}. Closed trajectories (C3) surrounding the center correspond to periodic solutions, while heteroclinic (C4) connections  between the saddle points indicate the existence of solitary wave solutions. The analytical solution presented in Eq.~\eqref{sec}, expressed in terms of the secant function, represents periodic wave structures, which agrees with the phase portrait analysis.

\paragraph{Case 2: $A_4>0,\;A_2>0$.}
The origin is the only equilibrium point and behaves as a saddle point. As shown in Fig.~\ref{d_b}, no closed trajectories exist, indicating the absence of periodic solutions.
The analytical solution given in Eq.~\eqref{csch}, expressed in terms of the hyperbolic cosecant function, corresponds to this parameter regime and describes a singular solitary-type wave behavior. This prediction is consistent with the analytical solution, which represents a singular and nonlocalized wave profile, confirming the absence of stable localized structures.
\paragraph{Case 3: $A_4<0,\;A_2<0$.}
The system possesses a single equilibrium point $(0,0)$, which is a center. All trajectories are closed curves; therefore, only periodic solutions arise. The phase-plane structure indicates the existence of periodic traveling-wave solutions in the regime $A_4<0,\;A_2<0$. However, these solutions are not captured by the present analytical reduction, because the corresponding quadrature is not real in this case. See Fig.~\ref{d_c}.

\paragraph{Case 4: $A_4<0,\;A_2>0$.}
In this situation, the equilibrium points consist of one saddle at the origin and two centers at
\[
\left(\pm\sqrt{-\frac{A_2}{2 A_4}},0\right).
\]
The phase portraits shown in Fig.~\ref{d_d} contain closed trajectories (C8) around the centers as well as separatrix structures, indicating the coexistence of periodic and solitary-type solutions. The analytical solution given in Eq.~\eqref{sech}, expressed in terms of the hyperbolic secant  function, corresponds to a localized solitary wave, confirming the theoretical prediction and representing a homoclinic orbit that acts as a separatrix between qualitatively different dynamical behaviors.

\begin{remark} The comparison between phase portrait predictions and analytical solutions shows complete agreement. In particular:
\begin{itemize}
\item Periodic solutions arise in regimes where center equilibria exist.
\item Singular or nonlocalized solutions appear when the system possesses only saddle equilibria.
\item Localized solitary wave solutions arise in the presence of saddle connections.
\end{itemize}
\end{remark}

 When $\alpha\neq 0$ in \eqref{dinamikgecis}, the quadratic nonlinearity breaks the $U\mapsto -U$ symmetry of the reduced traveling-wave equation. As a consequence, the phase portraits become asymmetric, the equilibrium configuration undergoes an imperfect (symmetry-breaking) bifurcation, and the corresponding exact traveling-wave solutions are generally expressed in terms of Jacobi elliptic functions.

 \section{Conclusion}

In this study, the $(2+1)$-dimensional Boussinesq equation with an arbitrary nonlinear term $f(u)$ was investigated by means of Lie symmetry analysis, reductions to ordinary differential equations, exact traveling-wave solutions, and stability analysis of the reduced traveling-wave system. Keeping $f(u)$ arbitrary, rather than restricting it to a prescribed polynomial form from the outset, makes the model more general and reveals a richer symmetry structure.

The symmetry classification shows that the equation admits a four-dimensional Lie algebra for arbitrary $f(u)$. In addition, the determining equations yield four distinct admissible forms of the nonlinear term \(f(u)\), and these forms produce an additional infinitesimal generator, extending the symmetry algebra to five dimensions. This shows that the symmetry structure of the equation is directly determined by the form of the nonlinearity.

These symmetries and corresponding transformations are related to the physical symmetries of the problem under consideration. The nonlinearity $f(u)$ is also related to the physical properties of the medium in which the waves propagate. Once it is determined which forms of this arbitrary function admit which symmetries, the corresponding forms of the nonlinearity that allow the model to possess symmetries corresponding to physical aspects of the problem under consideration can also be identified. The admissible forms of the nonlinearity are not merely algebraic outcomes of the symmetry analysis, but they may also indicate physically relevant structures of the propagation medium. In this way, results obtained through a purely theoretical analysis can be connected to concrete physical applications. For example, in \cite{genros25}, the authors performed a similar classification analysis for the Rosenau equation with a general nonlinearity. They obtained exponential and logarithmic functional forms for the admissible cases of the nonlinearity, and in Section 1.1 of their article, they derived a Rosenau equation with an exponential nonlinearity by following the same procedure as in \cite{rosenau88}. This is a good example of how a theoretical approach can lead to physically meaningful outcomes. In our case, the resulting equations \eqref{eqexp} and \eqref{eqlog}  are worth considering from a similar point of view, and we expect that they will attract the interest of researchers. Further, although Eq. \eqref{eqexp} seems to  lack the $u_{xx}$ term essential to the structure of the Boussinesq type equations, the series expansion of the exponential term provides such a term.  

In addition to the symmetry classification, we performed a systematic reduction analysis for the canonical equations obtained in the classification. Using an optimal system of two-dimensional subalgebras, we derived the corresponding similarity variables and reduced ordinary differential equations. These reductions were presented in Tables~1--5 for the arbitrary, exponential, logarithmic, power-law, and quadratic cases of the nonlinearity. In some cases, the reduced equations could be solved explicitly, and the corresponding invariant solutions were also provided. These reduced equations may serve as a useful starting point for further analytical, qualitative, and numerical studies of the canonical models.

For the exact solution analysis, the nonlinear form \(f(u)=\alpha u^2+\beta u^3\) was considered. Under the traveling-wave transformation, the equation was reduced to an ordinary differential equation and exact solutions were obtained. When both quadratic and cubic nonlinear effects are present, the solutions are expressed in terms of Jacobi elliptic functions. In certain limiting cases, these solutions reduce to simpler hyperbolic and trigonometric wave forms, showing that the model admits localized, singular, and periodic traveling-wave structures.

Furthermore, in order to study the stability and qualitative behavior of these traveling-wave solutions, the reduced ordinary differential equation was rewritten as a planar dynamical system. The resulting phase portraits are in agreement with the wave profiles obtained from the traveling-wave analysis and clarify the occurrence of periodic, solitary-type, and singular solutions in different parameter regimes.\\

\section*{Acknowledgements}
This work was supported by Scientific Research Projects Department of Istanbul Technical University, Project Number: TDK-2026-48158.



\end{document}